\documentclass[fleqn,12pt,twoside]{article}
\usepackage{espcrc1}

\input epsf

\usepackage{graphicx}
\usepackage[figuresright]{rotating}

\title{Mid-rapidity $\pi^\pm$, $K^\pm$, and $\overline{p}$ spectra and particle ratios from STAR}

\author{Olga Barannikova\address[MCSD]{Department of Physics, Purdue University, West Lafayette, Indiana 47907, USA} 
and Fuqiang Wang\addressmark\ for the STAR Collaboration\footnote{For the full author list and acknowledgments, see Appendix ``Collaborations" of this volume.}}
       
\begin{document}

\maketitle

\begin{abstract}
Results are presented on $\pi^\pm$, $K^\pm$, and $\overline{p}$ transverse mass spectra and particle multiplicity ratios at mid-rapidity in Au+Au collisions at $\sqrt{s_{_{\rm NN}}}$=130 and 200~GeV and in p+p collisions at $\sqrt{s}$=200~GeV. Comparisons are made to results from lower energies. The bulk properties of the collision inferred from these results are discussed.
\end{abstract}

\bigskip

New states in which the nucleon mass vanishes or quarks and gluons are deconfined have been predicted to exist at high densities over an extended volume~\cite{Lee,Shuryak}. Such states may be created in relativistic heavy-ion collisions~\cite{Lee,Shuryak}. Pions, kaons, and antiprotons are abundantly produced particles in relativistic heavy-ion collisions, reflecting the bulk properties of the collision: the yield of pions reflects on total entropy; kaons carry a significant fraction of the total strangeness produced; and the antiproton yield is a measure of baryon production. Measurements of these particles thus provide diagnostics to possible formation of the predicted new states~\cite{Harris,Bass}.

We present mid-rapidity transverse mass ($m_\perp$) spectra of charged pions ($\pi^\pm$), charged kaons ($K^\pm$), and antiproton ($\overline{p}$) in Au+Au collisions at RHIC. We study spectral shapes in the picture of collective transverse radial flow and production of kaons and antiprotons relative to that of pions. We investigate the systematics of these results with respect to the collision centrality and in the context of results from lower energies to search for systematic changes which could result from possible changes in the collision dynamics.

The data presented here are from minimum bias Au+Au collisions at nucleon-nucleon center-of-mass energy of $\sqrt{s_{_{\rm NN}}}$=130 and 200~GeV and p+p collisions at $\sqrt{s}$=200~GeV by the STAR experiment~\cite{Ackermann}. Tracks were reconstructed in the STAR Time Projection Chamber (TPC). The magnetic field was 0.25 and 0.5 Tesla for the 130 and 200~GeV data, respectively. The primary vertex of the interaction was found by fitting the tracks to a common point of origin. Tracks used in the analysis were required to come from within 3~cm of the primary vertex. Corrections were made for the energy loss of particles in the detector material. Particles were identified by measuring the specific energy loss ($dE/dx$) of charged particles in the TPC gas. The $dE/dx$ resolution was estimated to be 11\% which imposed upper momentum limits on the identified particle spectra. Corrections were applied to account for losses due to limited acceptance, decays, tracking inefficiencies, and hadronic interactions. 
The centrality bins used in the analysis were obtained by subdividing the measured mid-rapidity charged particle multiplicity in the TPC~\cite{pbarPaper,kaonPaper,Calderon,Cardenas}.

\begin{figure}[htb]
\centerline{
%
\epsfxsize=0.28\textwidth\epsfbox[110 240 430 580]{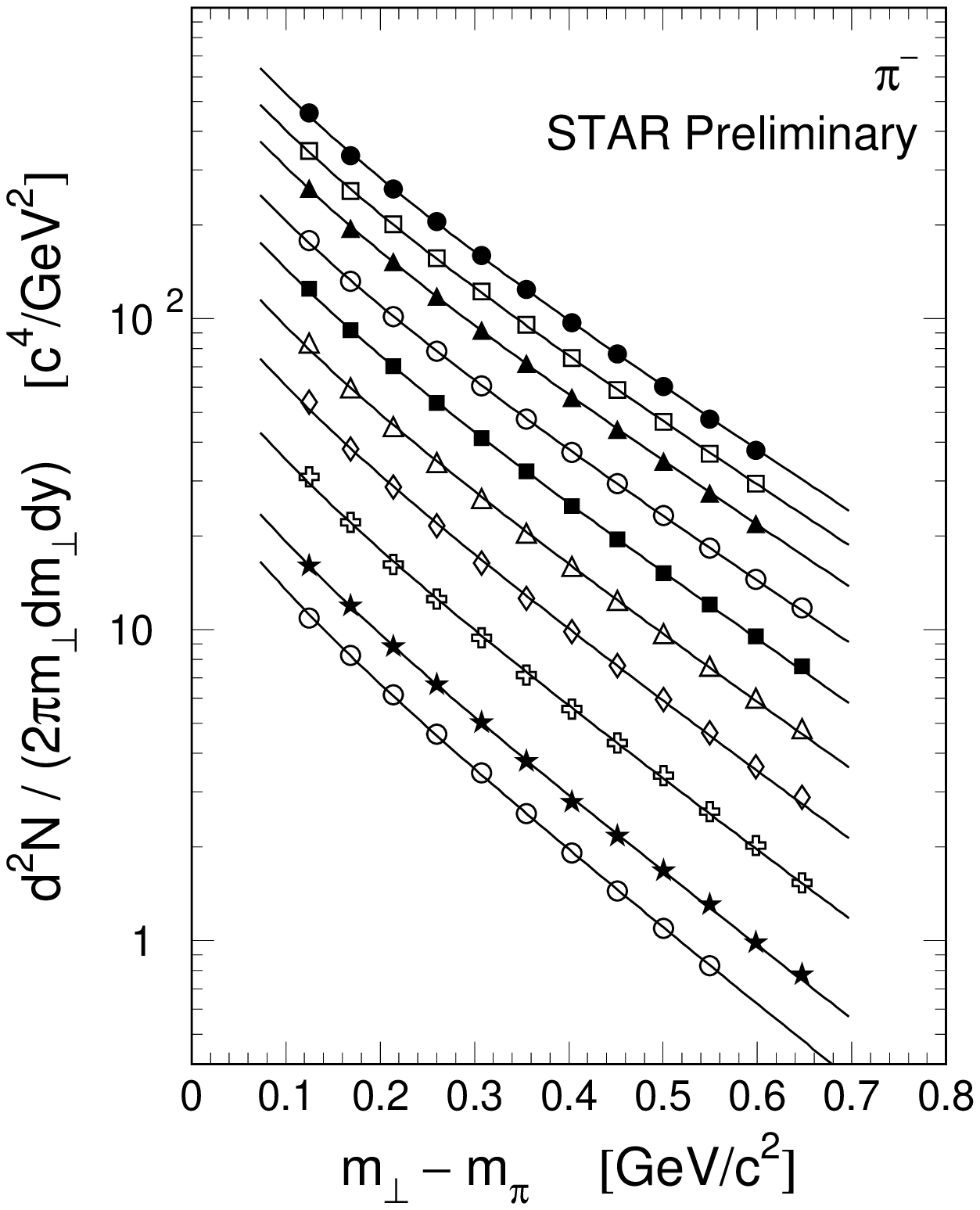}
\epsfxsize=0.28\textwidth\epsfbox[110 240 430 580]{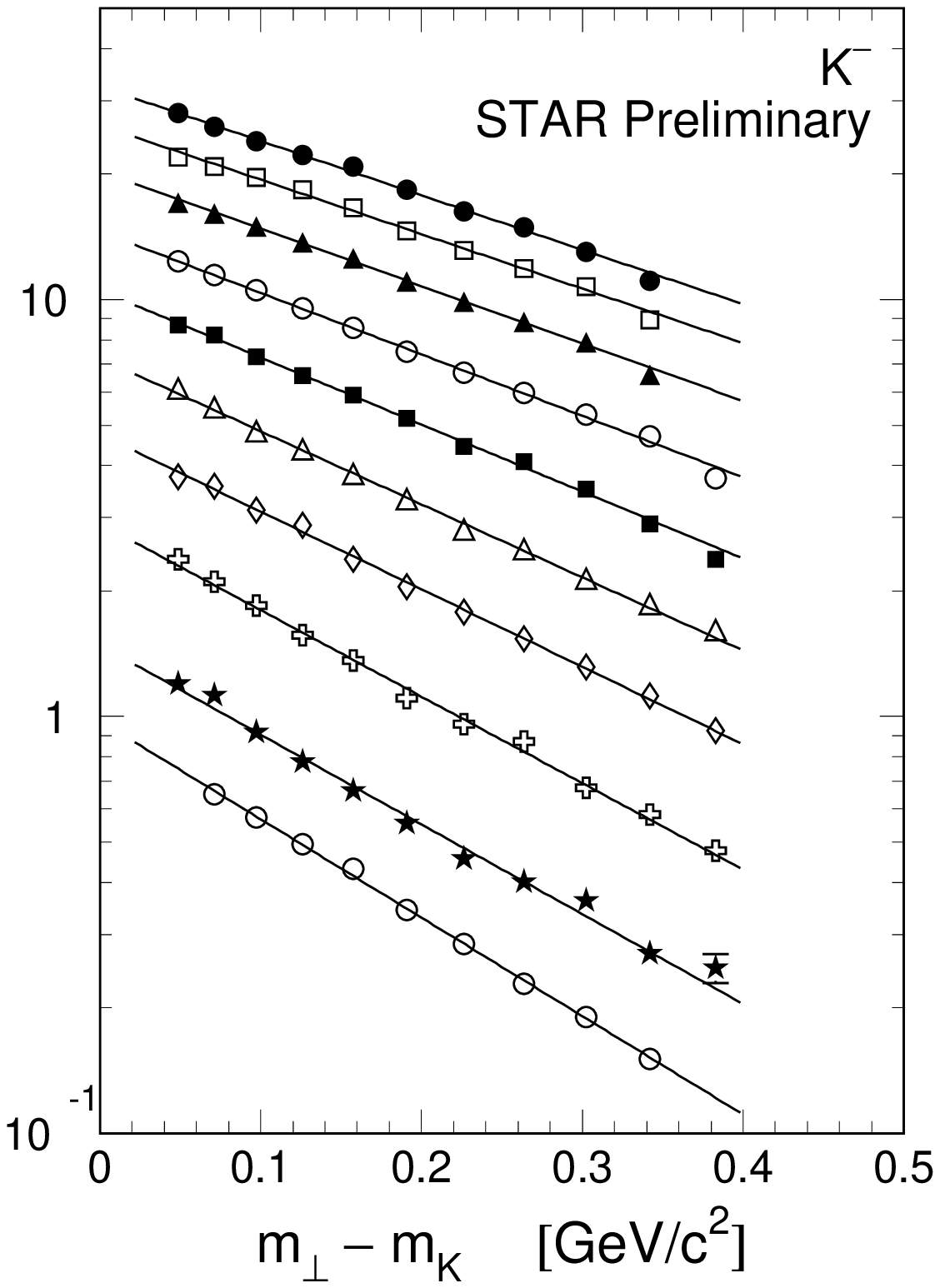}
\epsfxsize=0.28\textwidth\epsfbox[110 240 430 580]{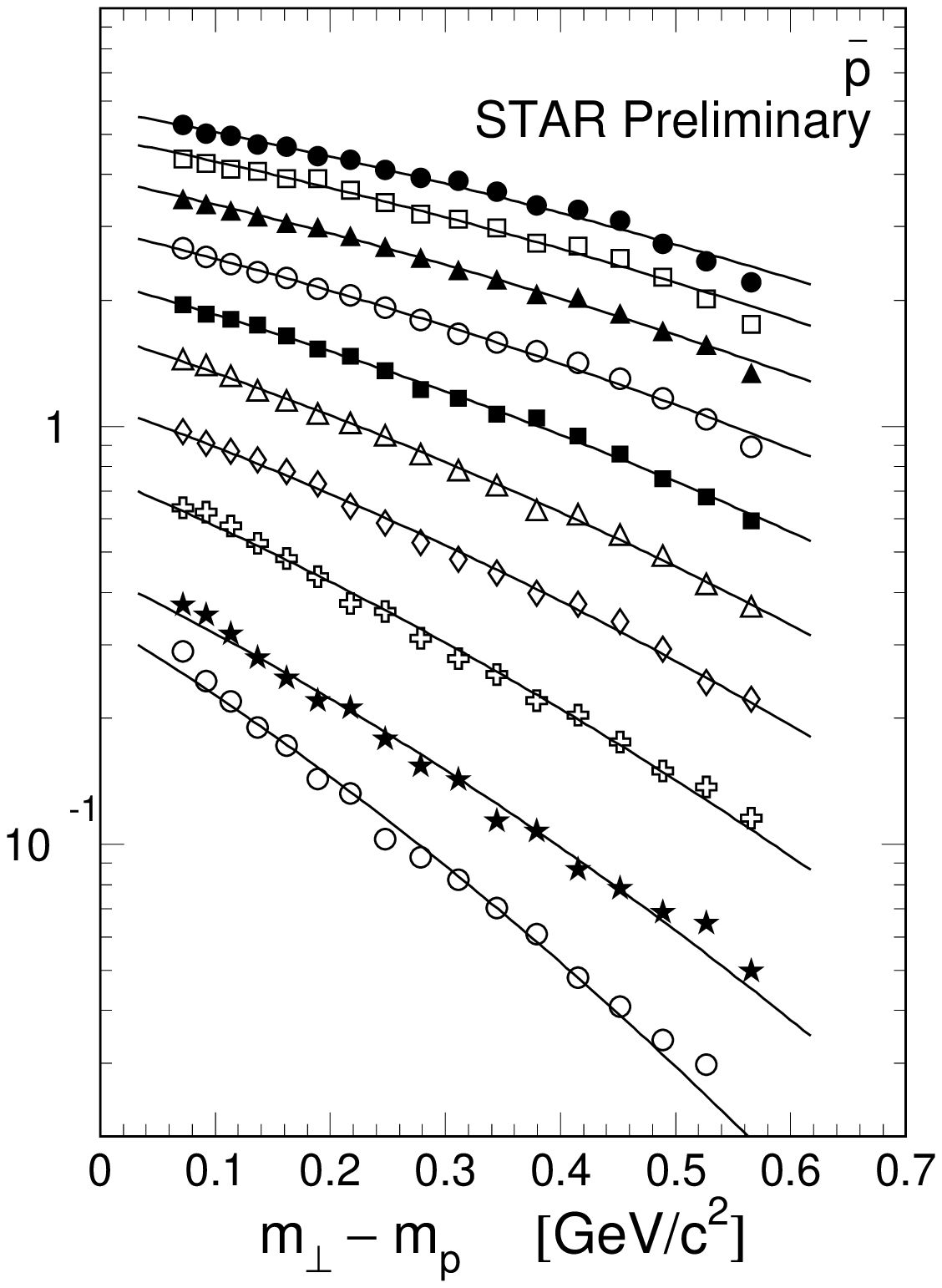}
}
%
%
\caption{Preliminary mid-rapidity transverse mass spectra of $\pi^-$ (left), $K^-$ (middle), and $\overline{p}$ (right). The lowest spectrum is from 200 GeV p+p within $|{\rm y}|<0.25$ (scaled up by a factor of 5). The other spectra are from 200 GeV Au+Au within $|{\rm y}|<0.1$, in the order of decreasing centrality from top to bottom: 5\% (most central), 5-10\%, 10-20\%, 20-30\%, 30-40\%, 40-50\%, 50-60\%, 60-70\%, and 70-80\%. The pion spectra are corrected for weak decays and muon contaminations. The kaon and antiproton spectra are inclusive. Statistical error bars are shown or smaller than the symbol size. See text for systematic errors and descriptions of the solid curves.}
\label{fqwang_fig1}
\end{figure}

Figure~\ref{fqwang_fig1} shows the mid-rapidity $\pi^-$, $K^-$, and $\overline{p}$ spectra for all centrality bins in 200~GeV Au+Au collisions. Systematic errors on the spectra are estimated to be 10\%. The $\pi^+$ and $K^+$ spectra are similar to the $\pi^-$ and $K^-$ spectra, respectively. All the spectra are similar to the corresponding ones at 130~GeV~\cite{pbarPaper,kaonPaper,Calderon,Cardenas}. We fit the pion spectra with a Bose-Einstein function, the kaon spectra with a $m_\perp$ exponential function, and the antiproton spectra with a $p_\perp$ gaussian function, to extract the mean transverse momenta, $\langle p_\perp \rangle$, and the $dN/d{\rm y}$ yields. The fit results are superimposed in Fig.~\ref{fqwang_fig1}. The measured yields are about 65\% of the extrapolated $dN/d{\rm y}$ for pions, 50-65\% for kaons, and 50-75\% for antiproton, respectively. Similar to~\cite{pbarPaper}, we use other functional forms to estimate the systematic errors on the extrapolated $\langle p_\perp \rangle$ and $dN/d{\rm y}$. The total systematic errors are 10\% on $\langle p_\perp \rangle$, and 15\% on pion and kaon $dN/d{\rm y}$ and 15-25\% on antiproton $dN/d{\rm y}$.

Figure~\ref{fqwang_fig1} also shows the mid-rapidity $\pi^-$, $K^-$, and $\overline{p}$ spectra for minimum bias p+p collisions at 200~GeV. Primary vertex finding inefficiency, becoming more significant for lower multiplicity events, is not corrected for and contribute to the majority of the estimated systematic error of $^{+10}_{-35}$\% on the spectra. Same functional forms (as for the Au+Au data) are used to fit the p+p spectra to extract the $\langle p_\perp \rangle$ and $dN/d{\rm y}$. Possible systematic effects of the vertex finding inefficiency on $\langle p_\perp \rangle$ are investigated by studying multiplicity dependence of the spectra. The total systematic errors on $\langle p_\perp \rangle$ and $dN/d{\rm y}$ are estimated to be 10\% and $^{+15}_{-40}$\%, respectively.

%
%

\begin{figure}[htb]
\centerline{
\epsfxsize=0.45\textwidth\epsfbox[20 300 520 540]{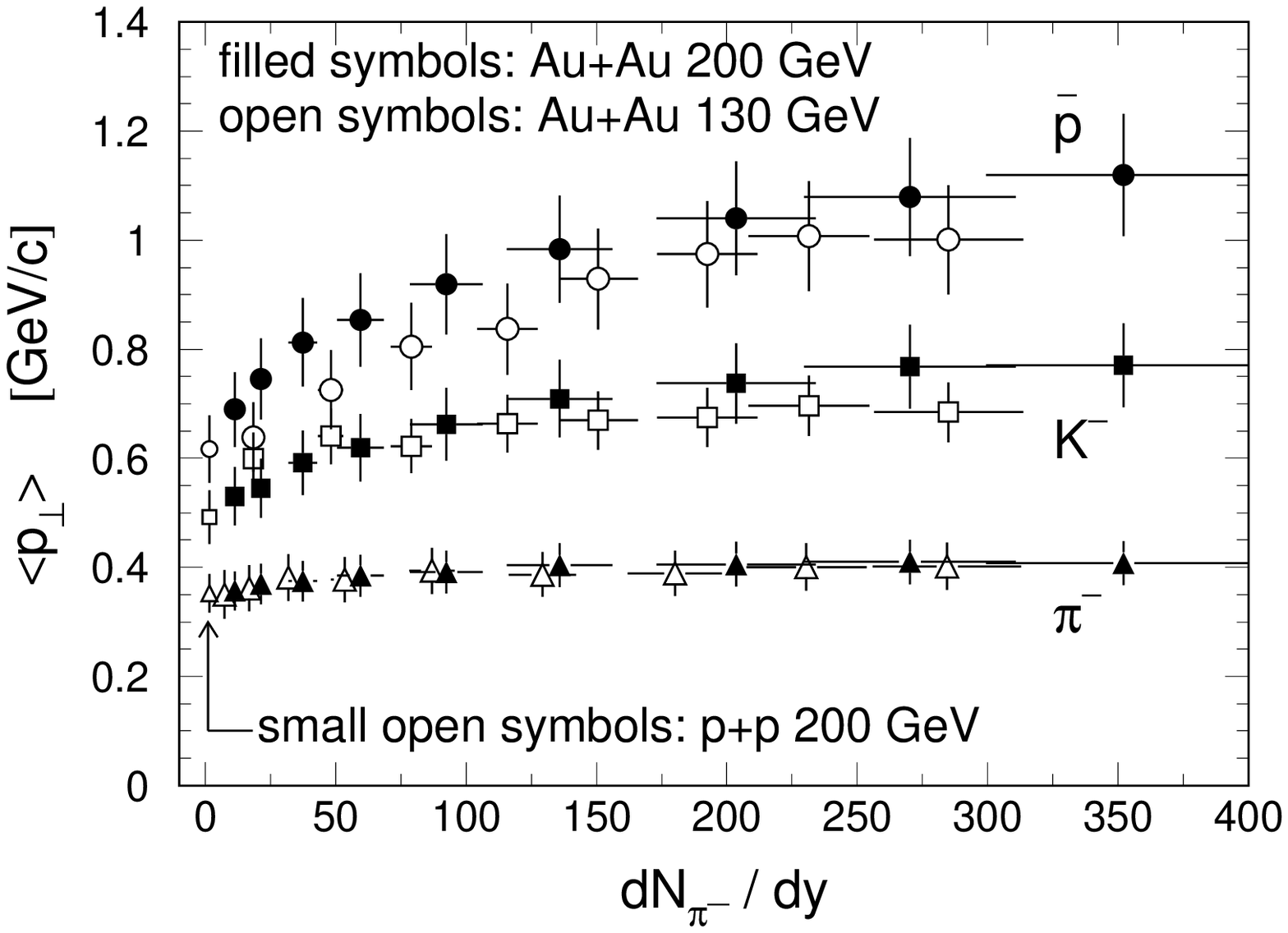}
\epsfxsize=0.45\textwidth\epsfbox[20 165 520 405]{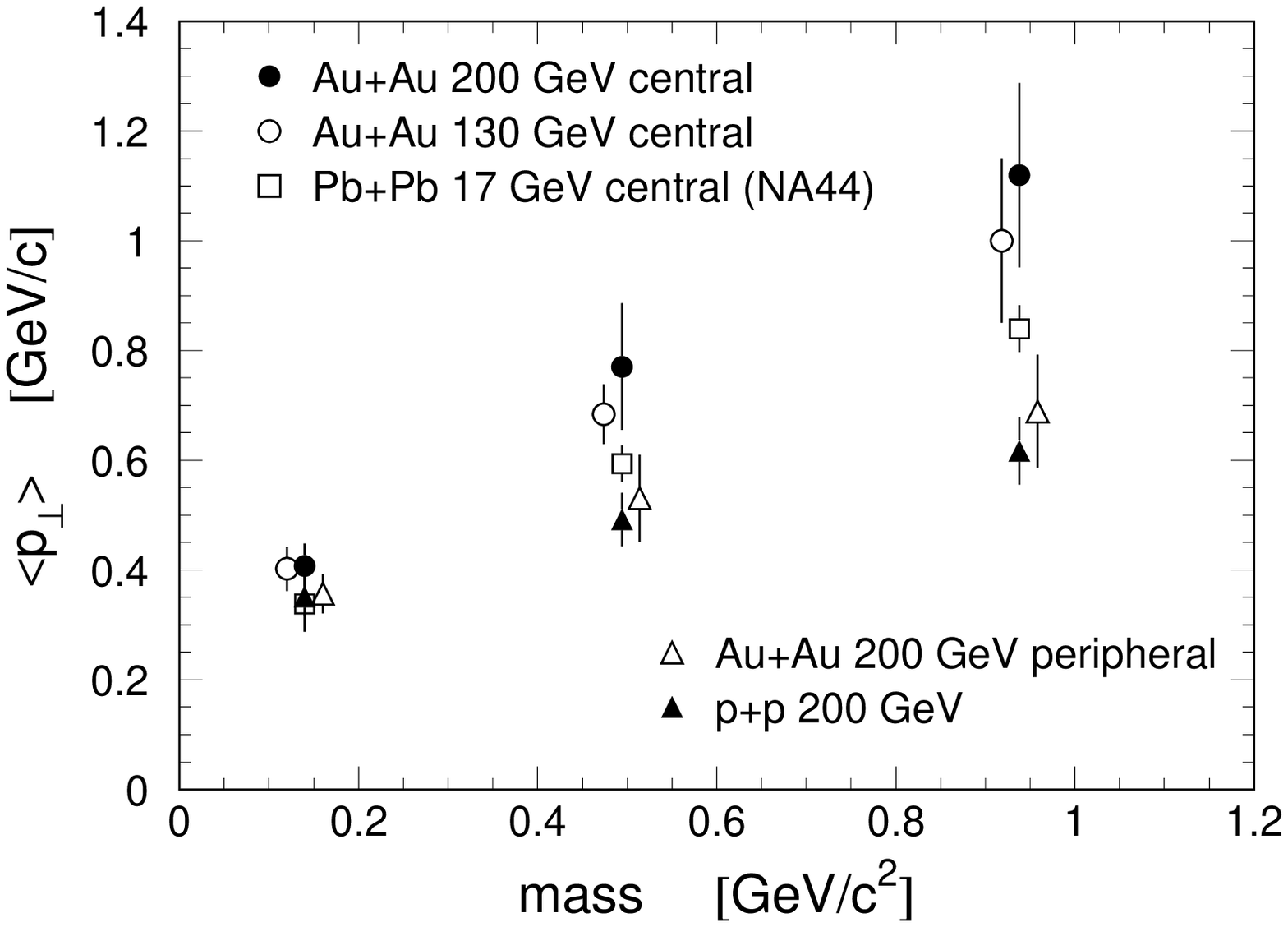}
}
\caption{Left: Extracted $\langle p_\perp \rangle$ versus centrality for $\pi^-$ (triangles), $K^-$ (squares), and $\overline{p}$ (circles) from STAR. Right: Extracted $\langle p_\perp \rangle$ versus particle mass for five collision systems. Errors shown for the STAR data are dominantly systematic, and for the SPS data~\cite{Bearden}, only statistical. The 200 GeV data and the 130 GeV pion data are preliminary.}
\label{fqwang_fig2}
\end{figure}

Figure~\ref{fqwang_fig2} (left) shows the extracted $\langle p_\perp \rangle$ of $\pi^-$, $K^-$, and $\overline{p}$ as a function of the mid-rapidity $\pi^-$ multiplicity density, $dN_{\pi^-}/d{\rm y}$, used as a measure of the collision centrality. A systematic increase with centrality is observed in the kaon and antiproton $\langle p_\perp \rangle$, consistent with collective transverse radial flow being built up in non-peripheral collisions. No significant change in $\langle p_\perp \rangle$ is observed from 130 to 200~GeV. Figure~\ref{fqwang_fig2} (right) shows $\langle p_\perp \rangle$ versus particle mass for five selected systems. The p+p and most peripheral A+A results indicate no transverse radial flow; the increase in $\langle p_\perp \rangle$ with mass in these systems is due to a trivial mass effect. The central collision results deviate from p+p and are consistent with transverse radial flow, which appears to be stronger at RHIC than SPS. Transverse radial flow is a property inferred from particle spectral shapes at kinetic freeze-out. It remains a question at what stage of the collision this collective flow is built up.

\begin{figure}[htb]
\centerline{
\epsfxsize=0.45\textwidth\epsfbox[0 150 500 440]{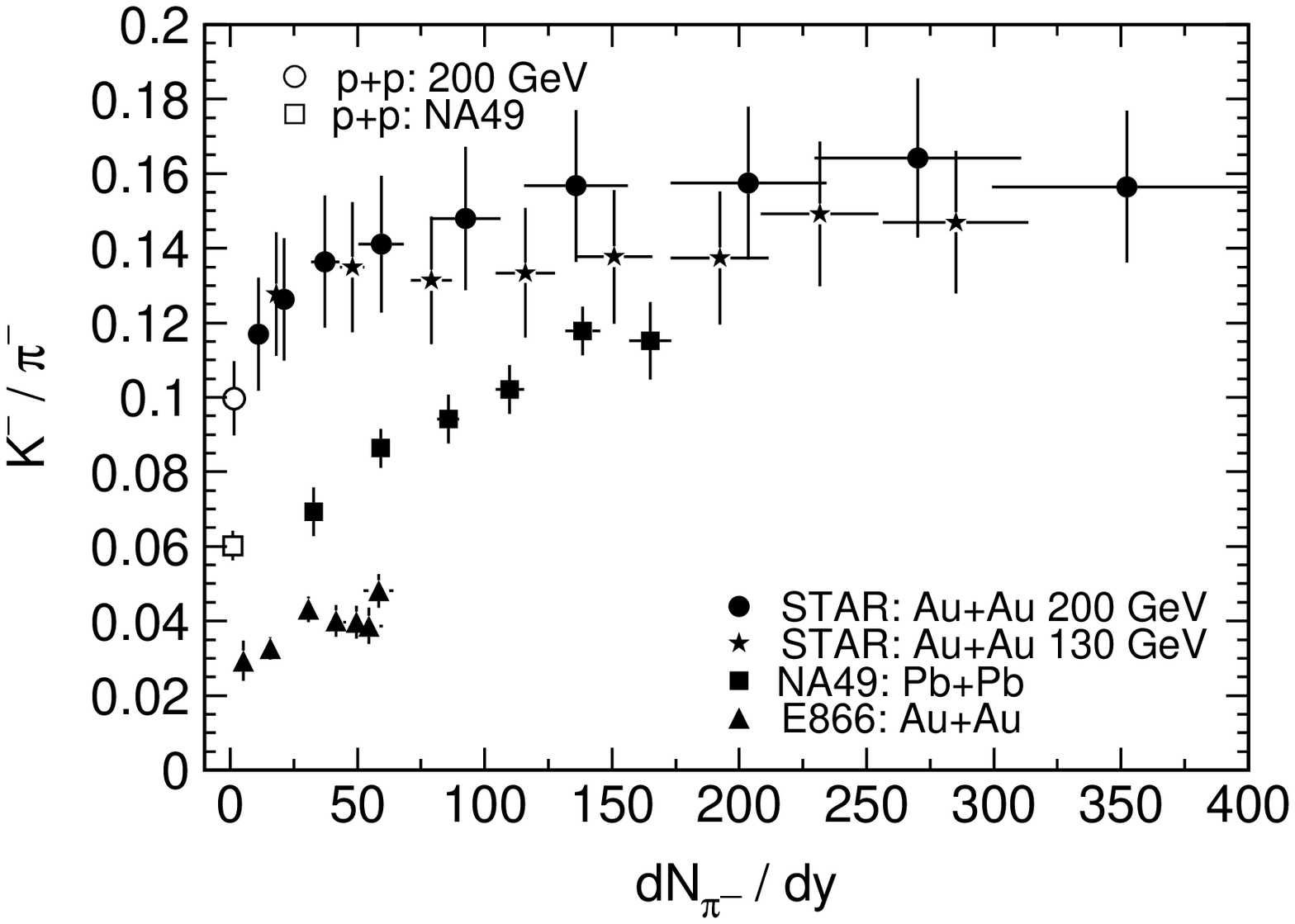}
\epsfxsize=0.45\textwidth\epsfbox[0 150 500 440]{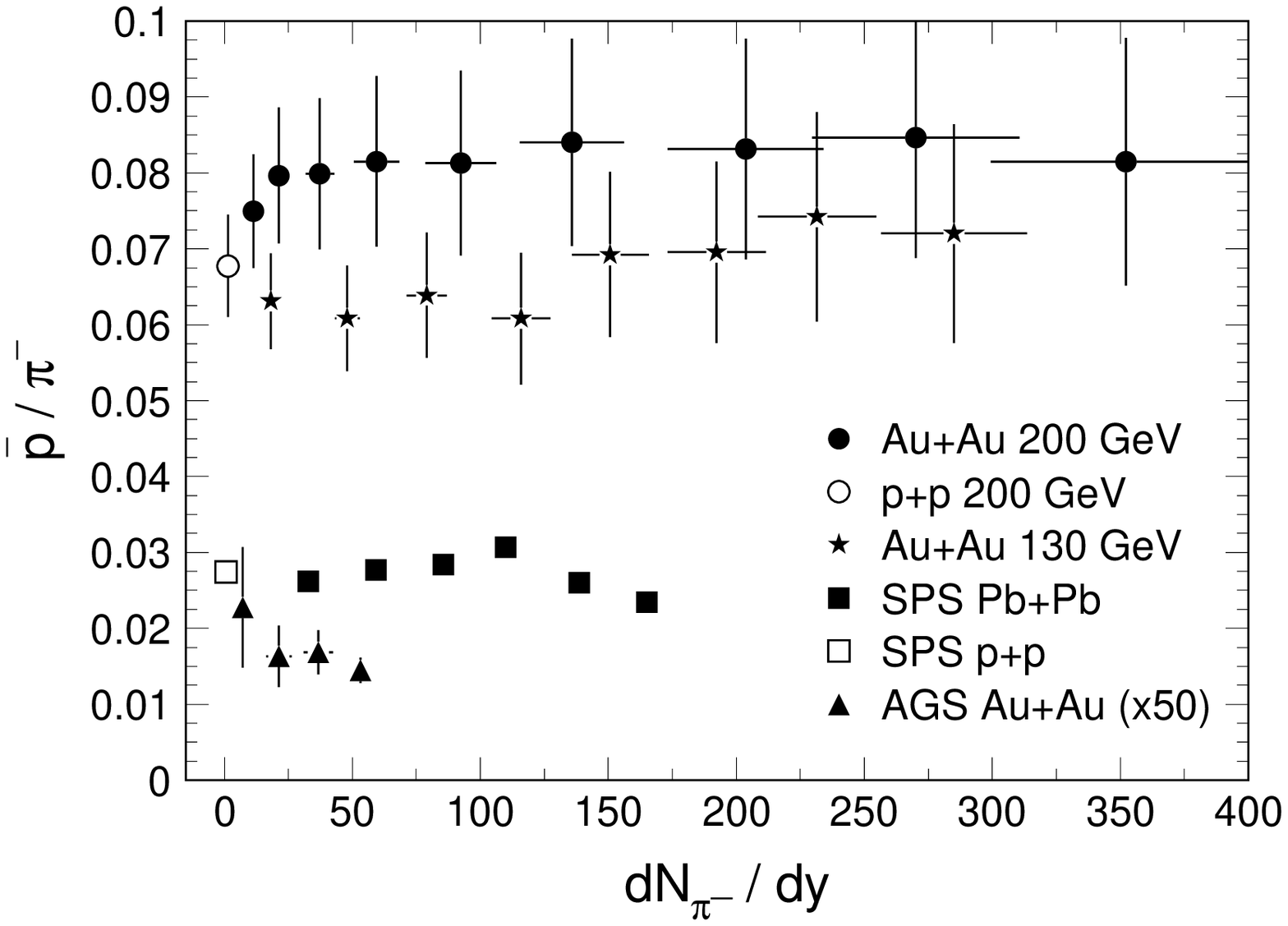}
}
\caption{Mid-rapidity $K^-/\pi^-$ (left) and $\overline{p}/\pi^-$ (right) ratios as a function of mid-rapidity $dN_{\pi^-}/dy$ from AGS/E866~\cite{E866kaon,E866pbar}, SPS/NA49~\cite{Sikler}, and RHIC/STAR (preliminary). Errors shown for the STAR data are dominantly systematic, and for the AGS~\cite{E866kaon,E866pbar} and SPS data~\cite{Sikler}, only statistical.}
\label{fqwang_fig3}
\end{figure}

Figure~\ref{fqwang_fig3} (left) compares the mid-rapidity $K^-/\pi^-$ ratio as a function of centrality at RHIC to those at the AGS~\cite{E866kaon} and SPS~\cite{Sikler}. The p+p results are also shown. At RHIC energies, little centrality dependence is found for $K^-/\pi^-$ once a certain centrality is reached. The increase from p+p to A+A is achieved in peripheral collisions. This behavior was not seen in lower energy collisions. Both the AGS and SPS data dispatch a linear increase of $K^-/\pi^-$ with $dN_{\pi^-}/dy$, suggesting that the physics might be different at RHIC than at lower energies. It has been suggested that the longitudinal thickness of the colliding nuclei may play a significant role at low energies and become less so at RHIC for medium-central to central collisions~\cite{Wang,Hohne}. Low energy RHIC data, expected in the coming years, should shed more light onto the physics.

Figure~\ref{fqwang_fig3} (right) shows the $\overline{p}/\pi^-$ ratio as a function of centrality at AGS~\cite{E866pbar}, SPS~\cite{Sikler}, and RHIC. Similar to $K^-/\pi^-$, the $\overline{p}/\pi^-$ ratio at RHIC is constant over centrality from medium-central to central collisions. Unlike $K^-/\pi^-$, however, the $\overline{p}/\pi^-$ ratios at the AGS and SPS are also flat, or perhaps decrease, with centrality~\cite{E866pbar,Sikler}. This might be related to antiproton absorption at later stages of the collision at the low energies. On the other hand, due to the relatively large production threshold, there is a dramatic increase in antiproton production from AGS to SPS to RHIC. 

In summary, we have presented results from STAR on mid-rapidity $\pi^\pm$, $K^\pm$, and $\overline{p}$ transverse mass spectra and $K^-/\pi^-$ and $\overline{p}/\pi^-$ multiplicity ratios in Au+Au collisions at 130 and 200~GeV and p+p collisions at 200~GeV. No significant changes were observed from 130 to 200~GeV, and there is a smooth evolution from p+p to A+A collisions. The following bulk properties of the collision may be inferred from these results: (1) At kinetic freeze-out, significant transverse radial flow has developed in central Au+Au collisions. The magnitude of radial flow increases with the collision centrality and is larger at RHIC than SPS. (2) At chemical freeze-out, the $K^-/\pi^-$ and $\overline{p}/\pi^-$ multiplicity ratios are higher in A+A than in p+p, with the increase achieved quickly but smoothly from p+p to peripheral A+A collisions, and remains relatively constant in medium-central to central collisions. This behavior is distinctly different from those observed at the AGS and SPS, suggesting a possible change in the production mechanisms of kaons and antiprotons with respect to pions from low energies to central collisions at RHIC energies. The present data may put stringent constraints on and provide inputs to the modeling of relativistic heavy-ion collisions.

\end{document}